\documentstyle[preprint,aps,epsf,floats]{revtex}

\begin{document}

\preprint{\tighten \vbox{\hbox{} }}

\title{Bounds on Heavy-to-Heavy Baryonic Form Factors}
\author{Cheng-Wei Chiang\footnote{\tt chengwei@andrew.cmu.edu}}

\address{
Department of Physics,
Carnegie Mellon University,
Pittsburgh, PA 15213}

\maketitle

{\tighten
\begin{abstract}
Upper and lower bounds are established on the $\Lambda_b \rightarrow
\Lambda_c$ semileptonic decay form factors by utilizing inclusive
heavy-quark-effective-theory sum rules.  These bounds are calculated
to leading order in $\Lambda_{\rm QCD}/m_Q$ and $\alpha_s$.  The
$O(\alpha_s^2\beta_0)$ corrections to the bounds at zero recoil are
also presented.  Several form factor models used in the literature are
compared with our bounds.
\end{abstract}
}%end tighten

%%\pacs{}

\newpage

\section{Introduction}
In the early development of heavy quark physics, much attention were
put on the semileptonic heavy meson decays $B \rightarrow D^{(*)} l
\bar\nu$ in order to extract information about the Kobayashi-Maskawa
matrix element $|V_{cb}|$ \cite{N91}.  As more and more data will
accumulate for semileptonic heavy baryon decays, they can also serve
as another independent determination of $|V_{cb}|$
\cite{CDF96,OPAL97,BL97}.

We have recently performed a thorough analysis of the bounds on the $B
\rightarrow D^{(*)}$ weak decay form factors using inclusive sum rules
for semileptonic decays to order $\Lambda_{\rm QCD}/m_Q$ and first
order in $\alpha_s$ \cite{CL99}.  The ${\mathcal O}(\alpha_s^2
\beta_0)$ corrections at zero recoil were also included.  In the
present paper, we would like to extend the techniques in our previous
work to baryons that contain a heavy quark.

As in the case of heavy mesons, heavy-to-heavy baryonic form factors
are mainly taken from models.  They are extensively used in the
studies of backgrounds and efficiencies in experiments, and, for this
reason alone, constraining them is important.  For $\Lambda_b
\rightarrow \Lambda_c$ transitions, where the initial and final
baryons have the same light degrees of freedom, the Heavy Quark
Effective Theory (HQET) relates all six form factors to a single
Isgur-Wise function and predicts a specific value at zero recoil in
the infinite mass limit \cite{IW91}.  In the HQET, the form factors and
Isgur-Wise function are usually written as functions of $\omega = v
\cdot v^{\prime}$, with $v$ being the four-velocity of the $\Lambda_b$
baryon and $v^{\prime}$ that of the recoiling $\Lambda_c$ baryon.

The analysis for heavy baryons is almost parallel to that for heavy
mesons.  One difference is the HQET parameters and heavy hadron masses
appearing in each case.  Another difference is that heavy baryons have
an extra light quark.  However, we will see that the power of
inclusive sum rules still applies, regardless of the intricacy among
the light degrees of freedom.

The layout of this paper is as follows: In Section II, we list the sum
rule formulae for the model-independent bounds on form factors defined
in Section III.  Section III also gives the proper combinations of
structure functions used later for the bounds on each of the baryonic
form factors.  Section IV provides the bounds on individual form
factors explicitly, with the structure functions given in \cite{CL99},
and discusses the influence of the various parameters appearing in the
expansion of the bounds.  Some popular form factor models are compared
with our bounds in Section V.  Order $\alpha_S^2 \beta_0$ corrections
to the bounds at zero recoil are computed in Section VI.  In Section
VII, the bounds on the $\Lambda_b \rightarrow \Lambda_c l \nu$ decay
spectrum and the slope are given.  Our conclusions are summarized in
Section VIII.

% ==============================================================

\section{Inclusive HQET sum rules}

The sum rules are obtained by relating the inclusive decay rate,
calculated using the operator product expansion (OPE) and perturbative
QCD in the partonic picture, to the sum of exclusive decays rates,
calculated in the hadronic picture.  Since these had been derived
previously\cite{BSUV95,BR97,BLRW97,CL99}, we just write down the
results below.  The same formalism can be applied to all types of
heavy baryons.  Nevertheless, we will concentrate exclusively on
$\Lambda$-type ground-state baryons and denote them by $\Lambda_Q$.

Let $\left|{\Lambda_Q}(v,s) \right>$ be the state of the heavy baryon
$\Lambda_Q$ with mass $M_{\Lambda_Q}$ and spin $s$ moving with
four-velocity $v$ and
$\left|{\Lambda_{Q^{\prime}}}(v^{\prime},s^{\prime}) \right>$ that of
the $\Lambda_Q^{\prime}$ with energy $E_{\Lambda_{Q^{\prime}}}$ and
spin $s^{\prime}$ moving with four-velocity $v^{\prime}$.
$E_{\Lambda_{Q^{\prime}}} = \sqrt{M_{\Lambda_{Q^{\prime}}}^2+q_3^2}$,
where $q_3$ is the $z$ component of the momentum transfer, $q$, to the
leptonic sector and $\vec{q}$ is pointing in the $z$ direction.
Consider the time ordered product of two weak transition currents
between $\Lambda_Q$ baryons in momentum space,
\begin{eqnarray}
\label{structurefn}
T^{\mu\nu}& = & \frac{i}{2M_{\Lambda_Q}} \frac1{2j+1} \sum_{s}
   \int d^4x \, e^{-iq \cdot x} 
   \left< \Lambda_Q(v,s) \right| T(J^{\mu\dagger}(x)J^{\nu}(0)) 
   \left| \Lambda_Q(v,s) \right> 
   \nonumber \\
& = &-g^{\mu\nu}T_1 + v^{\mu}v^{\nu}T_2 
   + i \epsilon^{\mu\nu\alpha\beta}q_{\alpha}v_{\beta}T_3 
   + q^{\mu}q^{\nu}T_4 + (q^{\mu}v^{\nu}+v^{\mu}q^{\nu})T_5,
\end{eqnarray}
where $J^\mu$ is a $Q \rightarrow Q^{\prime}$ axial or vector current
and $j$ is the spin of $\left|{\Lambda_Q}(v,s) \right>$.. The bounds
are then
\begin{eqnarray}
\label{uplowb}
\frac{1}{2\pi i} \int_{C} d\epsilon \, \theta(\Delta-\epsilon) \,
   T(\epsilon) \left( 1-\frac{\epsilon}{E_1-E_{\Lambda_{Q^{\prime}}}} \right) 
&\leq&
   \frac{M_{\Lambda_{Q^{\prime}}}}{2j+1} \sum_{s,s^{\prime}}
   \frac{\left| \left< {\Lambda_{Q^{\prime}}}(v^{\prime},s^{\prime}) 
   \right| a \cdot J \left|
   {\Lambda_Q}(v,s) \right> \right|^2}{4M_{\Lambda_Q} E_{\Lambda_{Q^{\prime}}}}
   \nonumber \\ 
&\leq& \; \frac{1}{2\pi i} \int_{C} d\epsilon \,
   \theta(\Delta-\epsilon) \, T(\epsilon),
\end{eqnarray}
where $T(\epsilon) \equiv a_{\mu}^* T^{\mu \nu} a_{\nu}$, the product
of $T^{\mu \nu}$ and four-vectors $a_{\mu}^* a_{\nu}$.  The
integration variable $\epsilon = M_{\Lambda_Q} -
E_{\Lambda_{Q^{\prime}}} - v \cdot q$, $E_1$ is the energy of the
first excited state more massive than $\Lambda_{Q^{\prime}}$, and
$\Delta$ is the scale up to which the perturbation sums over.

The upper bound is essentially model independent, while the lower
bound relies on the assumption that there is no multiparticle
production of hadrons in the final state.  This additional assumption
is in accord with the large $N_c$ limit and is supported as well by
current experimental results.  These bounds can be used for the decays
at arbitrary momentum transfer $q^2$.  Since
$1/(E_1-E_{\Lambda_{Q^{\prime}}}) \sim 1/\Lambda_{\rm QCD}$, the lower
bounds will be limited to first order in $1/m_Q$.  The upper bounds,
on the other hand, can be calculated to order $1/m_Q^2$ without
additional HQET parameters.

The above formulae are presented in terms of $\Lambda$-type baryons,
however, they can be readily applied to other types of baryons too.
Later on, we will restrict our analysis exclusively to the
$\Lambda$-type semileptonic decays with $Q \rightarrow b$ and
${Q^{\prime}} \rightarrow c$, {\it i.e.}, the $\Lambda_b \rightarrow
\Lambda_c l \bar\nu$ decays for which $j=1/2$.

% ==============================================================

\section{Hadronic side}

The hadronic matrix elements for semileptonic decays of a $\Lambda_b$
baryon into a $\Lambda_c$ baryon are conventionally parameterized in
terms of six form factors defined by
\begin{eqnarray}
\label{param}
\left< \Lambda_c(v^\prime,s^\prime)\mid V^\mu \mid \Lambda_b(v,s)\right>
  & = &
  {\bar u}_{\Lambda_c}(v^\prime,s^\prime) 
  \left[ F_1(\omega) \gamma^\mu + F_2(\omega) v^\mu 
       + F_3(\omega) {v^\prime}^\mu \right]
  u_{\Lambda_b}(v,s), \nonumber \\
\\ 
\left< \Lambda_c(v^\prime,s^\prime)\mid A^\mu \mid \Lambda_b(v,s)\right>
  & = &
  {\bar u}_{\Lambda_c}(v^\prime,s^\prime) 
  \left[ G_1(\omega) \gamma^\mu + G_2(\omega) v^\mu 
       + G_3(\omega) {v^\prime}^\mu \right] \gamma_5
  u_{\Lambda_b}(v,s). \nonumber
\end{eqnarray}
One may relate the parameter $\omega$ to the momentum transfer $q^2$
by $\omega = (M_{\Lambda_b}^2+M_{\Lambda_c}^2-q^2)/(2M_{\Lambda_b}
M_{\Lambda_c})$.  With proper choices of the current $J^{\mu}$ and the
four-vector $a_{\mu}$, one can readily select the form factor of
interest and, from Eq.~(\ref{uplowb}), form the corresponding bounds.
To subleading order in $1/m_Q$ in HQET, these form factors satisfy the
relations: \cite{GGW90,FN93}
\begin{eqnarray}
\label{hqetformfactorrelation}
 F_1 &=& \zeta(\omega) \left[
         1 + \bar\Lambda \left( \frac1{2m_c}+\frac1{2m_b} \right) \right], \\
 G_1 &=& \zeta(\omega) \left[
         1 + \bar\Lambda \left( \frac1{2m_c}+\frac1{2m_b} \right)
         \frac{\omega-1}{\omega+1} \right], \nonumber \\
 F_2(\omega) &=& G_2(\omega) =
 -\frac{\bar\Lambda}{m_c}\frac1{\omega+1} \zeta(\omega), \nonumber \\
 F_3(\omega) &=& -G_3(\omega) =
 -\frac{\bar\Lambda}{m_b}\frac1{\omega+1} \zeta(\omega). \nonumber
\end{eqnarray}
In the above equations, we have absorbed the corrections from the
subleading kinematic energy into the Isgur-Wise function,
$\zeta(\omega)$.

To bound $F_1$ (or $G_1$), one may choose $J^{\mu}=V^{\mu}$ (or
$A^{\mu}$) and $a^{\mu}=(0,1,0,0)$.  Then the factor to be bounded is
$\frac{(\omega-1)}{2\omega} \left| F_1(\omega) \right|^2 \; \left(
\frac{\omega+1}{2\omega} \left| G_1(\omega) \right|^2 \right)$ and the
sum rule used to bound is $T_{1OPE}=T_{1Hadronic}$.  In this and the
following cases, the corresponding first excited state more massive
than $\Lambda_c$ that contributes to the sum rule is the
$\Lambda_c(2593)$ state.\footnote{$\Sigma_c(2455)$ is lighter, but
$\Lambda_b \rightarrow \Sigma_c(2455) l \nu$ is not an allowed
transition because of the isospin.}

It is impossible to bound $F_2$ and $F_3$ individually with any choice
of $a^\mu$.  The next best thing we can do is to bound a linear
combination of them.  To prevent the bounds from diverging at zero
recoil, we choose $J^{\mu}=V^{\mu}$ and $a^\mu =
(\sqrt{E_{\Lambda_c}/M_{\Lambda_c}-1},0,0,-\sqrt{E_{\Lambda_c}/M_{\Lambda_c}-1})$.
Then the factor to be bounded is $\frac{\omega^2-1}{2\omega} \left|
F_2-F_3 \right|^2$, and the sum rule requires the following
combination of structure functions
\begin{eqnarray}
\label{F23sum}
T(\epsilon) &=& 
    2 T_1 + \left( \omega-1 \right) T_2 
  + \left( M_{\Lambda_b}+M_{\Lambda_c}-\epsilon \right)^2 
    \left( \omega-1 \right) T_4 \\
&&+ 2 \left( M_{\Lambda_b}+M_{\Lambda_c}-\epsilon \right)
      \left( \omega-1 \right) T_5. \nonumber
\end{eqnarray}

A similar situation occurs for $G_2$ and $G_3$.  Here the choice would
be $J^{\mu}=A^{\mu}$ and $a^\mu =
(1,0,0,-\sqrt{\frac{E_{\Lambda_c}-M_{\Lambda_c}}{E_{\Lambda_c}+M_{\Lambda_c}}})$.
The bounded factor is $\frac{\omega-1}{2\omega} \left| G_2+G_3
\right|^2$, and
\begin{eqnarray}
\label{G23sum}
T(\epsilon) &=& 
    -\frac{2}{\omega+1} T_1 + T_2 
  + \left( M_{\Lambda_b}-M_{\Lambda_c}-\epsilon \right)^2 T_4 \\
&&+ 2 \left( M_{\Lambda_b}-M_{\Lambda_c}-\epsilon \right) T_5 \nonumber
\end{eqnarray}
is the combination for the sum rule.

From Eq.~(\ref{hqetformfactorrelation}), we know
$|F_2-F_3|^2=|G_2+G_3|^2 \sim 1/m_Q^2$ at this order in HQET.

% ==============================================================

\section{Partonic side and the bounds}

The analysis will be performed to first order in $\alpha_s(m_Q)$
($\sim 0.3$ at $2 {\rm\ GeV}$) and $\Lambda_{\rm QCD}/m_Q$ and,
furthermore, only terms linear in $\omega-1$ are kept for the
perturbative part.  The full $\omega$ dependence will be kept in the
nonperturbative physics.  Corrections of order $\Lambda_{\rm
QCD}^2/m_Q^2$, $\alpha_s^2$, $\alpha_s\,\Lambda_{\rm QCD}/m_Q$, and
$\alpha_s\,(\omega -1)^2$ should be negligible in the kinematic region
that we are considering.

We will use essentially the same notation as in \cite{CL99}.  Since
$\Lambda_b$ and $\Lambda_c$ are spin-$\frac{1}{2}$ baryons where the
light degrees of freedom carry zero angular momentum, the
chromomagnetic energy $\lambda_2=0$.  Also, the heavy quark parameters
$\bar\Lambda$ and $\lambda_1$ for the baryons are different from those
for the mesons.  Their values, however, can be obtained from those of
mesons by the two relations \cite{MW94}
\begin{eqnarray}
\label{hqetparas}
\bar\Lambda^B &=& \bar\Lambda^M + \frac{m_b}{m_b-m_c} \left(
              M_{\Lambda_b}-{\bar M_B} \right)
              - \frac{m_c}{m_b-m_c} \left( M_{\Lambda_c}-{\bar M_D}
              \right), \nonumber \\
\lambda_1^B &=& \lambda_1^M
            + \frac{2 m_b m_c}{m_b-m_c} \left[ \left(
            M_{\Lambda_b}-{\bar M_B} \right) -
            \left( M_{\Lambda_c}-{\bar M_D} \right) \right], \\
{\bar M_B} &=& \frac{M_B+3M_{B^*}}{4},
\qquad
{\bar M_D} = \frac{M_D+3M_{D^*}}{4}. \nonumber
\end{eqnarray}
In Eq.~(\ref{hqetparas}), $\bar\Lambda^B$ and $\lambda_1^B$ are the
parameters for baryons and $\bar\Lambda^M$ and $\lambda_1^M$ those for
mesons.    Later on, we will go back to the usual notation without
attaching superscripts explicitly.

To form the bounds, one just takes the proper moments of the structure
functions according to the combination required in the sum rules given
in Section III.  To this order, the $\Lambda_{\rm QCD}/m_Q$ corrections will
depend on two HQET parameters; $\bar\Lambda$, $\lambda_1$.  As
mentioned above, we take their values from the mesonic ones, where a
certain linear relation has been determined \cite{GKLW96}.  We will
use three different parameter sets to show the dependence on
$\bar\Lambda$ and $\lambda_1$: (A) $\bar\Lambda = 0.74 {\rm\ GeV}$ and
$\lambda_1 = -0.43 {\rm\ GeV}^2$, (B) $\bar\Lambda = 0.64 {\rm\ GeV}$
and $\lambda_1 = -0.33 {\rm\ GeV}^2$, (C) $\bar\Lambda = 0.84 {\rm\
GeV}$ and $\lambda_1 = -0.53 {\rm\ GeV}^2$.

The sum rule for bounding $(\omega-1) \left| F_1(\omega)
\right|^2/(2\omega)$ uses $T_1$ with vector-vector currents.  The
upper bound is simply the zeroth moment of $T_1$, which is by
Eq.~(\ref{uplowb})
\begin{equation}
\label{F1upper}
\frac{(\omega-1)}{2\omega} \left| F_1(\omega) \right|^2 \leq
   I_1^{(0)VV} + A_1^{(0)VV}.
\end{equation}
The first moment of $T_1$ is needed for the lower bound,
which is
\begin{equation}
\label{F1lower}
\frac{(\omega-1)}{2\omega} \left| F_1(\omega) \right|^2 \geq
   I_1^{(0)VV} + A_1^{(0)VV}
- \frac{1}{E_{\Lambda_1}-E_{\Lambda_c}}
     \left( I_1^{(1)VV} + A_1^{(1)VV}\right)
\end{equation}
The upper and lower bounds are shown in Fig.~1.\footnote{For all the
figures in this section we take $m_b = 4.8 {\rm\ GeV}$, $m_c = 1.4
{\rm\ GeV}$, $\alpha_s = 0.3$ (corresponding to a scale of about $2
{\rm\ GeV}$).  The values of $\bar\Lambda$ and $\lambda_1$ are
discussed in the text.}
\begin{figure}[t]
\centerline{\epsfysize=11truecm  \epsfbox{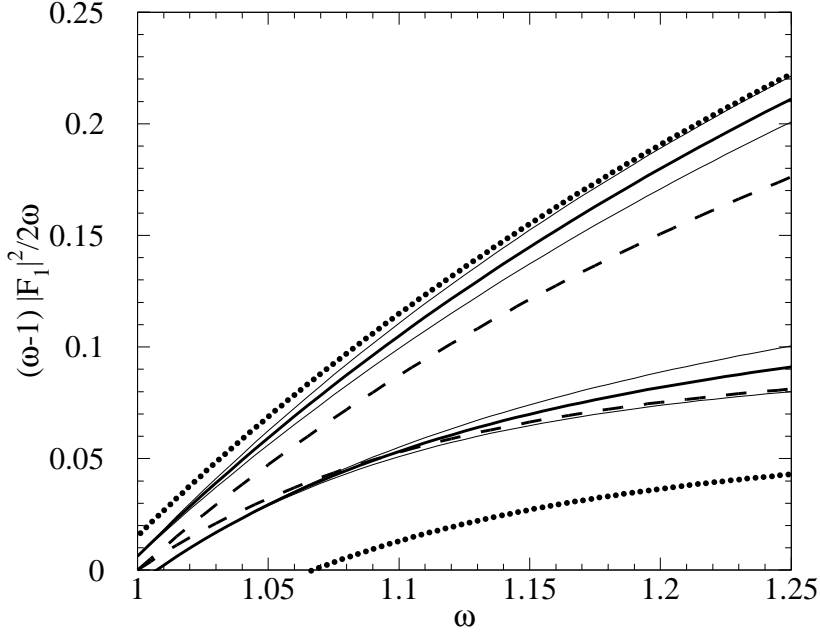} }
\tighten{
\caption[]{\it Upper and lower bounds on $(\omega-1) \left|
F_1(\omega) \right|^2/(2\omega)$.  The thick solid (dotted) curves
are the upper and lower bounds including perturbative corrections for
HQET parameter set (A) described in the text, and $\Delta = 1{\rm\ GeV}\
(2{\rm\ GeV})$.  The dashed curves are the bounds without perturbative
corrections, also for HQET parameter set (A).  The thin solid curves
show the dependence on $\bar\Lambda$ and $\lambda_1$, using parameter
sets (B) and (C), with $\Delta = 1{\rm\ GeV}$.  Here the outer two
thin solid curves are from parameter set (B), while the inner two
curves from set (C)}}
\end{figure}
For this section, the dotted curves are the bounds without
perturbative corrections using set (A) above, while the solid and
dashed curves represent the bounds including the perturbative
corrections with $\Delta=1 {\rm\ GeV}$ and $\Delta=2 {\rm\ GeV}$,
respectively.  We have shown the bounds in the kinematic range $1 \leq
\omega \lesssim 1.25$, where the higher order correction
$\alpha_s(\omega -1)^2$ should be negligible.  The thin solid curves
use the other HQET parameter sets (B) and (C).

In Fig.~1, the bounds without perturbative corrections approach zero
because of the overall factor $\omega-1$ appearing in the
nonperturbative corrections.  As we move the scale of $\Delta$ from $1
{\rm GeV}$ to $2 {\rm GeV}$, the upper bound is only shifted upward by
about $0.01$, while the lower bound is dragged down by more than
$0.04$.  This serves as an indication that in general perturbative
corrections have a larger influence on the lower bound.  The bounds
get wider at small momentum transfer with parameter set (B) while
narrower with set (C).  As a general tendency in this and subsequent
plots, varying among the three parameter sets (A), (B), and (C)
usually makes less change on the upper bound than the lower bound.
Here, the variations on the upper and lower bounds at $\omega=1$ are
about $5\%$ and $11\%$, respectively.

A similar set of bounds for $(\omega+1) \left| G_1(\omega)
\right|^2/2\omega$ can be obtained by simply changing $VV$
(vector-vector currents) to $AA$ (axial-axial currents) in the above
formulae.\footnote{This set of bounds has been mentioned in
\cite{BLRW97}, and we agree with their results.}  The bounds in this
case are shown in Fig.~2.  Perturbative corrections drag the bounds
lower.  From the bounds corresponding to $\Delta=2 {\rm GeV}$ we also
observe that the lower bound has been significantly decreased.  In
this case, the parameter set (B) lowers the bounding curves, but set
(C) pushes them up.  The variations of the upper and lower bounds at
$\omega=1$ by these parameter changes are about $1.5\%$ and $17\%$,
respectively.
\begin{figure}[t]
\centerline{\epsfysize=11truecm  \epsfbox{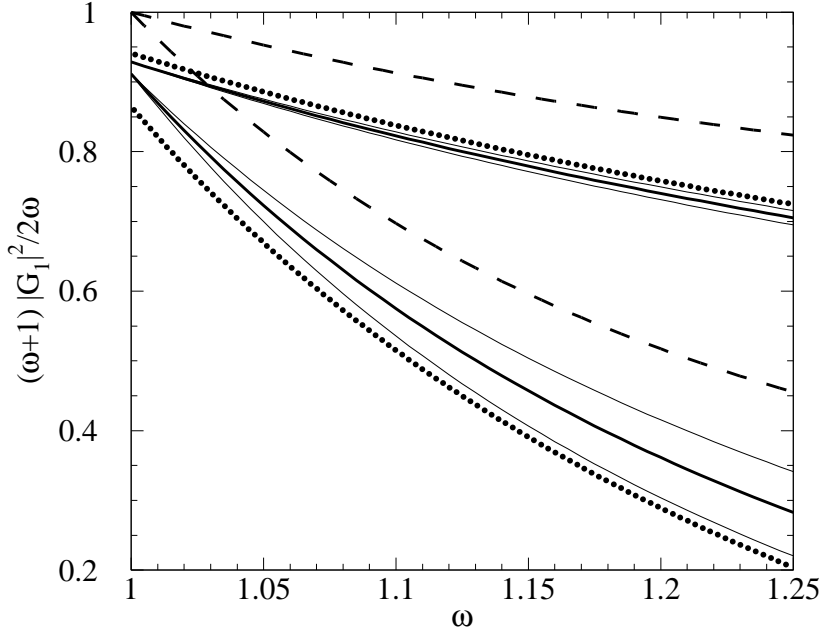} }
\tighten{
\caption[]{\it Upper and lower bounds on $(\omega+1) \left|
G_1(\omega) \right|^2/(2\omega)$.  The curves are labeled the same as
in Fig.~1.  Here the parameter set (B) gives the lower set of thin
solid curves while the set (C) gives the upper set of curves.}}
\end{figure}

The bounds on the other form factors involve higher moments of $T_4$
and $T_5$.  The upper bound for $(\omega^2-1) \left|
F_2(\omega)-F_3(\omega) \right|^2/2\omega$ is, from Eq.~(\ref{F23sum}),
\begin{eqnarray}
\label{F23upper}
F_{23}^{upper} &=& 
 (\omega-1) \left\{ \frac2{\omega-1} I_1^{(0)VV}
 + I_2^{(0)VV} 
 + I_4^{(2)VV}
 - 2 (M_{\Lambda_b}+M_{\Lambda_c}) I_4^{(1)VV} \right. \\
&&\phantom{(\omega-1)[]} 
 + (M_{\Lambda_b}+M_{\Lambda_c})^2 I_4^{(0)VV}
 - 2 I_5^{(1)VV}
 + 2 (M_{\Lambda_b}+M_{\Lambda_c}) I_5^{(0)VV}
\nonumber \\
&&\phantom{(\omega-1)[]}
 + \frac2{\omega-1} A_1^{(0)VV}
 + A_2^{(0)VV} 
 + A_4^{(2)VV}
 - 2 (M_{\Lambda_b}+M_{\Lambda_c}) A_4^{(1)VV} \nonumber\\
&&\phantom{(\omega-1)[]}
 \left. 
 + (M_{\Lambda_b}+M_{\Lambda_c})^2 A_4^{(0)VV}
 - 2 A_5^{(1)VV}
 + 2 (M_{\Lambda_b}+M_{\Lambda_c}) A_5^{(0)VV} \right\}.\nonumber
\end{eqnarray}
The lower bound is $(\omega^2-1) \left|
F_2(\omega)-F_3(\omega) \right|^2/2\omega \geq F_{23}^{lower}$, with
\begin{eqnarray}
\label{F23lower}
F_{23}^{lower} &=& F_{23}^{upper} \\
&&
 - \frac{(\omega-1)}{M_{\Lambda_1} - M_{\Lambda_c}} 
 \left\{
   \frac2{\omega-1} I_1^{(1)VV}
 + I_2^{(1)VV} 
 + I_4^{(3)VV}
 - 2 (M_{\Lambda_b}+M_{\Lambda_c}) I_4^{(2)VV} 
 \right. \nonumber\\
&& \phantom{- \frac{(\omega-1)}{M_{\Lambda_1} - M_{\Lambda_c}}}
 + (M_{\Lambda_b}+M_{\Lambda_c})^2 I_4^{(1)VV}
 - 2 I_5^{(2)VV}
 + 2 (M_{\Lambda_b}+M_{\Lambda_c}) I_5^{(1)VV}
\nonumber \\
&&\phantom{- \frac{(\omega-1)}{M_{\Lambda_1} - M_{\Lambda_c}}}
   \frac2{\omega-1} A_1^{(1)VV}
 + A_2^{(1)VV} 
 + A_4^{(3)VV}
 - 2 (M_{\Lambda_b}+M_{\Lambda_c}) A_4^{(2)VV} \nonumber\\
&&\phantom{- \frac{(\omega-1)}{M_{\Lambda_1} - M_{\Lambda_c}}}
 \left.
 + (M_{\Lambda_b}+M_{\Lambda_c})^2 A_4^{(1)VV}
 - 2 A_5^{(2)VV}
 + 2 (M_{\Lambda_b}+M_{\Lambda_c}) A_5^{(1)VV}
 \right\}.\nonumber
\end{eqnarray}

They are plotted in Fig.~3.  
\begin{figure}[t]
\centerline{\epsfysize=11truecm  \epsfbox{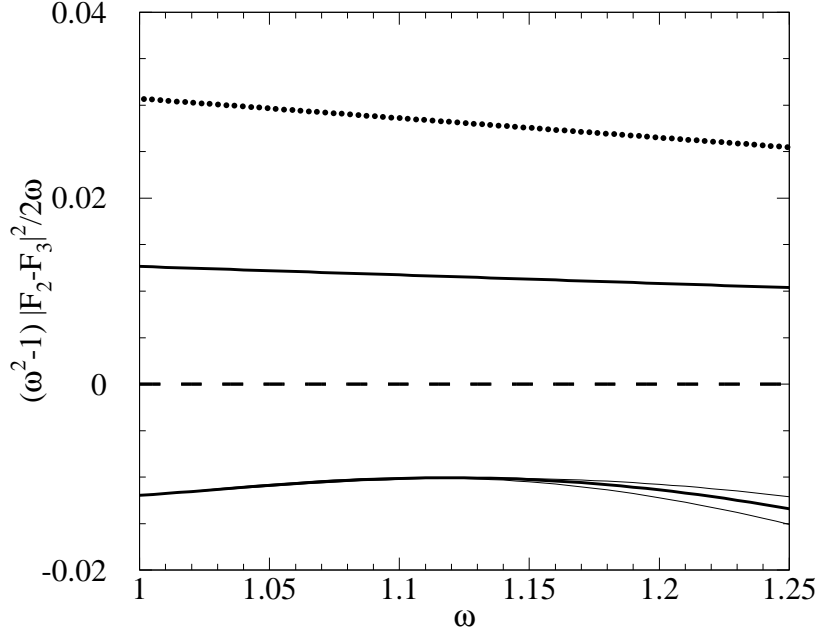} }
\tighten{
\caption[]{\it Upper and lower bounds on $(\omega^2-1) \left|
F_2(\omega) - F_3(\omega) \right|^2/2\omega$.  The curves are
labeled the same as in Fig.~1.}}
\end{figure}
Notice that at tree level both bounds are identically zero.  This can
be easily understood from the relations in
Eq.~(\ref{hqetformfactorrelation}), $\left| F_2-F_3 \right|^2 \sim
\bar\Lambda^2/m_Q^2$.  Nevertheless, perturbative physics separates
the bounds by pushing the upper bound higher to around 0.01 over the
entire kinematic range while drawing the lower bound negative.  Since
the factors we are bounding are all positive definite, the lower bound
is still zero.  Changing $\Delta$ from $1{\rm\ GeV}$ to $2{\rm\ GeV}$
loosens the upper bound by more than a factor of two.  Parameter sets
(A), (B), and (C) have no influence on the upper bound because
$\bar\Lambda$ and $\lambda_1$ only enter the upper bound through the
tree-level corrections and they vanish to the order under
consideration.

The upper bound for $(\omega-1) \left|
G_2(\omega)+G_3(\omega) \right|^2/2\omega$ is, from Eq.~(\ref{G23sum}),
\begin{eqnarray}
\label{G23upper}
G_{23}^{upper} &=& 
 - \frac2{\omega+1} I_1^{(0)AA}
 + I_2^{(0)AA} 
 + I_4^{(2)AA}
 - 2 (M_{\Lambda_b}-M_{\Lambda_c}) I_4^{(1)AA} \\
&&\qquad
 + (M_{\Lambda_b}-M_{\Lambda_c})^2 I_4^{(0)AA}
 - 2 I_5^{(1)AA}
 + 2 (M_{\Lambda_b}-M_{\Lambda_c}) I_5^{(0)AA}
\nonumber \\
&&\qquad
 + \frac2{\omega-1} A_1^{(0)AA}
 + A_2^{(0)AA} 
 + A_4^{(2)AA}
 - 2 (M_{\Lambda_b}-M_{\Lambda_c}) A_4^{(1)AA} \nonumber\\
&&\qquad
 + (M_{\Lambda_b}-M_{\Lambda_c})^2 A_4^{(0)AA}
 - 2 A_5^{(1)AA}
 + 2 (M_{\Lambda_b}-M_{\Lambda_c}) A_5^{(0)AA}. \nonumber
\end{eqnarray}
The lower bound is $(\omega-1) \left|
G_2(\omega)+G_3(\omega) \right|^2/2\omega \geq G_{23}^{lower}$, with
\begin{eqnarray}
\label{G23lower}
G_{23}^{lower} &=& G_{23}^{upper} \\
&&
 - \frac{1}{M_{\Lambda_1} - M_{\Lambda_c}} \left\{
 - \frac2{\omega+1} I_1^{(1)AA}
 + I_2^{(1)AA}
 + I_4^{(3)AA}
 - 2 (M_{\Lambda_b}-M_{\Lambda_c}) I_4^{(2)AA}
 \right. \nonumber\\
&& \phantom{- \frac{(\omega-1)}{M_{\Lambda_1} - M_{\Lambda_c}}}
 + (M_{\Lambda_b}-M_{\Lambda_c})^2 I_4^{(1)AA}
 - 2 I_5^{(2)AA}
 + 2 (M_{\Lambda_b}-M_{\Lambda_c}) I_5^{(1)AA}
\nonumber \\
&&\phantom{- \frac{(\omega-1)}{M_{\Lambda_1} - M_{\Lambda_c}}}
 - \frac2{\omega+1} A_1^{(1)AA}
 + A_2^{(1)AA}
 + A_4^{(3)AA}
 - 2 (M_{\Lambda_b}-M_{\Lambda_c}) A_4^{(2)AA} \nonumber\\
&&\phantom{- \frac{(\omega-1)}{M_{\Lambda_1} - M_{\Lambda_c}}}
 \left.
 + (M_{\Lambda_b}-M_{\Lambda_c})^2 A_4^{(1)AA}
 - 2 A_5^{(2)AA}
 + 2 (M_{\Lambda_b}-M_{\Lambda_c}) A_5^{(1)AA}
 \right\}.\nonumber
\end{eqnarray}

The bounds on this form factor are shown in Fig.~4.
\begin{figure}[t]
\centerline{\epsfysize=11truecm  \epsfbox{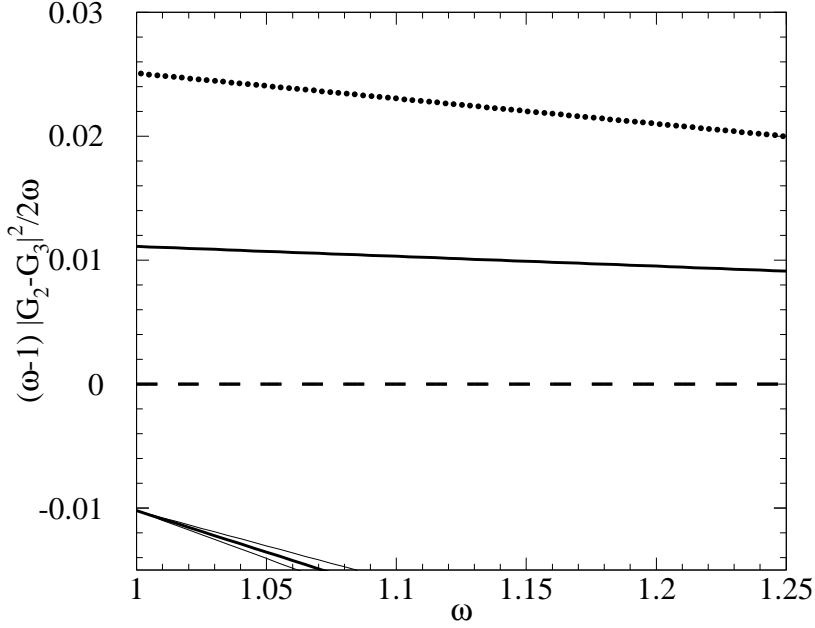} }
\tighten{
\caption[]{\it Upper and lower bounds on $(\omega-1) \left|
G_2(\omega) + G_3(\omega) \right|^2/2\omega$.  The curves are
labeled the same as in Fig.~1.}}
\end{figure}
Again, for the same reason as in the previous case, both the upper
and lower bounds at tree level are identically zero, and the upper
bound does not split for different parameter sets.  The perturbative
corrections also push both bounds in the current case away from zero;
and using $\Delta=2{\rm\ GeV}$ in the calculation widens the upper
bound by more than a factor of 2.

At ${\mathcal O}(\Lambda_{\rm QCD}/m_Q)$, the upper bounds will not
depend upon $\lambda_1$.  However, since $1/(M_{D1}-M_D) \sim
1/\Lambda_{\rm QCD}$ its value will affect the lower bounds.  It is
possible to obtain the upper bounds to order ${\mathcal
O}(\Lambda_{\rm QCD}^2/m_Q^2)$, since at this order there are no new
parameters in the OPE.  These corrections only slightly modify the
upper bounds in Figs.~1 and 2 but greatly loosen the ones in Figs.~3
and 4 to around 0.1 over the entire kinematic range.

% ==============================================================

\section{Comparison with models}

We choose from the literature the following commonly used form factor
models for comparison with our bounds:
\begin{verse}
(1) Soliton Model \cite{JMW92}, \\
(2) MIT Bag Model \cite{SZ93}, \\
(3) BS Model \cite{IKLR99}, \\
(4) Relativistic 3-Quark Model \cite{IKLR98}, \\
(5) COQM Model \cite{MGKIIO99}. \\
\end{verse}

The soliton model \cite{JMW92} considers baryons with a single heavy
quark $c$ (or $b$) in the large $N_c$ limit as bound states of $D$ and
$D^*$ mesons (or $B$ and $B^*$ mesons) with baryons containing only
light $u$ and $d$ quarks, which can be viewed as solitons in this
limit.  The MIT Bag Model used in \cite{SZ93} is actually a modified
one that takes into account the familiar picture from the
nonrelativistic quark model that treats the light component of a
baryon as a diquark.  The Relativistic 3-Quark Model \cite{IKLR98}
modifies the interaction Lagrangian between heavy hadrons and quarks
by incorporating the spin structure imposed by the spectator quark
model.  Finally, the COQM model \cite{MGKIIO99} uses the covariant
quark model.  All these models give Isgur-Wise functions without a
dependence on the heavy quark masses.  Thus, we use
Eqs.~(\ref{hqetformfactorrelation}) to relate the form factors to this
Isgur-Wise function.

Figs.~5-8 show the different form factors from the models and the
bounding curves.  In plotting these figures, we used $m_b = 4.8 {\rm\
GeV}$, $m_c = 1.4 {\rm\ GeV}$, $\alpha_s = 0.3$ (corresponding to a
scale of about $2 {\rm\ GeV}$), $\Delta = 1 {\rm\ GeV}$ and current
PDG data on heavy meson masses.  As mentioned above, $\lambda_1$ and
$\bar\Lambda$ are not easy to obtain experimentally.  Here we pick the
parameter set (A) discussed above, $\bar\Lambda = 0.74 {\rm\ GeV}$,
$\lambda_1 = -0.43 {\rm\ GeV}^2$.  The uncertainty on $\lambda_1$ and
$\bar\Lambda$ will correspondingly slightly modify our bounds, as
discussed before.

Fig.~5 shows the model values of $(\omega-1) \left|F_1(\omega)
\right|^2/(2\omega)$ along with the corresponding bounds for
comparison.  
\begin{figure}[t]
\centerline{\epsfysize=11truecm  \epsfbox{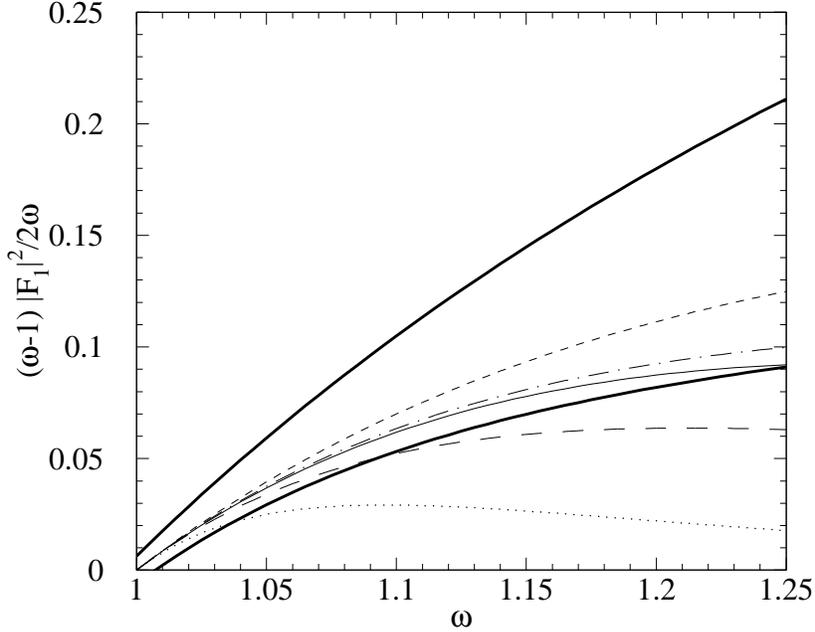} }
\tighten{
\caption[]{\it The model values of $(\omega-1) \left|F_1(\omega)
\right|^2/(2\omega)$ along with the corresponding bounds for
comparison.  The thick solid lines are the upper and lower bounds.
The thin solid curve is the Soliton Model.  The long dashed curve is
the MIT Model, and the short dashed curve is the BS Model.  The
dot-dashed curve is the Relativistic 3-Quark Model.  The dotted curve
is the COQM Model.}}
\end{figure}
The Soliton Model, Relativistic 3-Quark Model and the BS Model agree
with our bounds over the entire kinematic regime.  The MIT Bag Model
only slightly violates the lower bound, whereas the COQM model falls
far below the lower bound at large recoil.  If a larger $\Delta$ value
is employed, all the model curves would fit into the bounds as one can
check from Fig.~1.

The curves for
$(\omega+1)\left| G_1 \right|^2/2\omega$ are shown in Fig.~6.  
\begin{figure}[t]
\centerline{\epsfysize=11truecm  \epsfbox{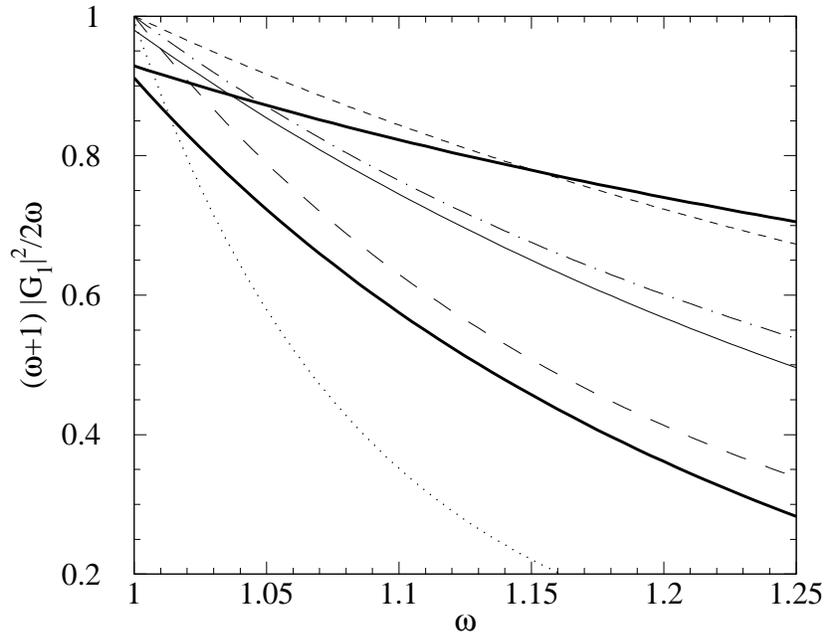} }
\tighten{
\caption[]{\it The model values of $(\omega+1)\left| G_1 \right|^2/2\omega$
along with the corresponding bounds for comparison.  The curves are
labeled the same as in Fig.~5.}}
\end{figure}
All models start from one at zero recoil.  This is because they do not
incorporate the perturbative renormalization effect.  The COQM Model
quickly becomes too small in the large recoil regime.  This, along
with Fig.~5, tells us that the contribution from $\Lambda_b
\rightarrow \Lambda_c$ at small momentum transfer is underestimated in
the COQM Model.  The other models are consistent with our bounds at
large recoil and, presumably, will nicely fit into our bounds near
zero recoil if renormalization corrections are included.

Fig.~7 and Fig.~8 show that all the model predictions are well within
our bounds for $(\omega^2-1)\left| F_2-F_3 \right|^2/(2\omega)$ and
$(\omega-1)\left| G_2+G_3 \right|^2/2\omega$, respectively.
\begin{figure}[t]
\centerline{\epsfysize=11truecm  \epsfbox{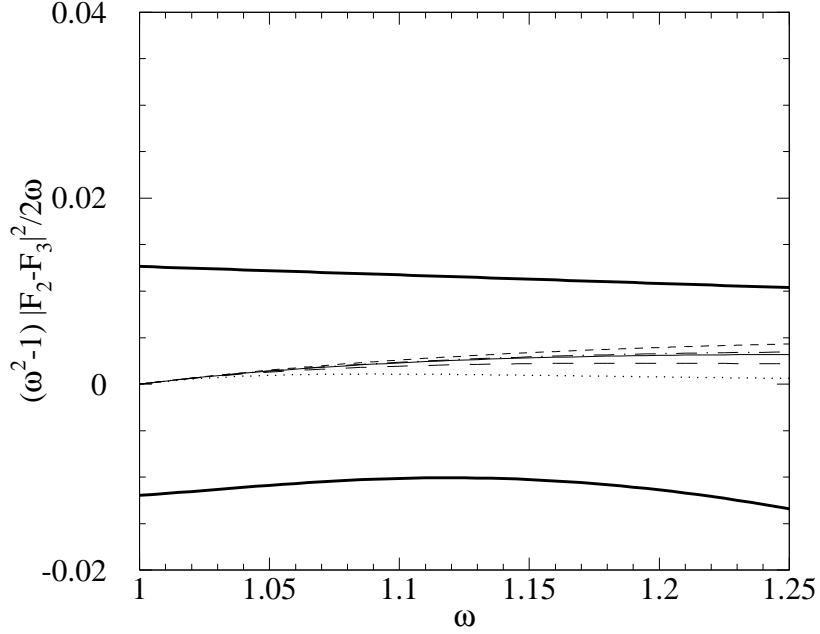} }
\tighten{
\caption[]{\it The model values of $(\omega^2-1)\left| F_2-F_3
\right|^2/(2\omega)$ along with the corresponding bounds for
comparison.  The curves are labeled the same as in Fig.~5.}}
\end{figure}
\begin{figure}[t]
\centerline{\epsfysize=11truecm  \epsfbox{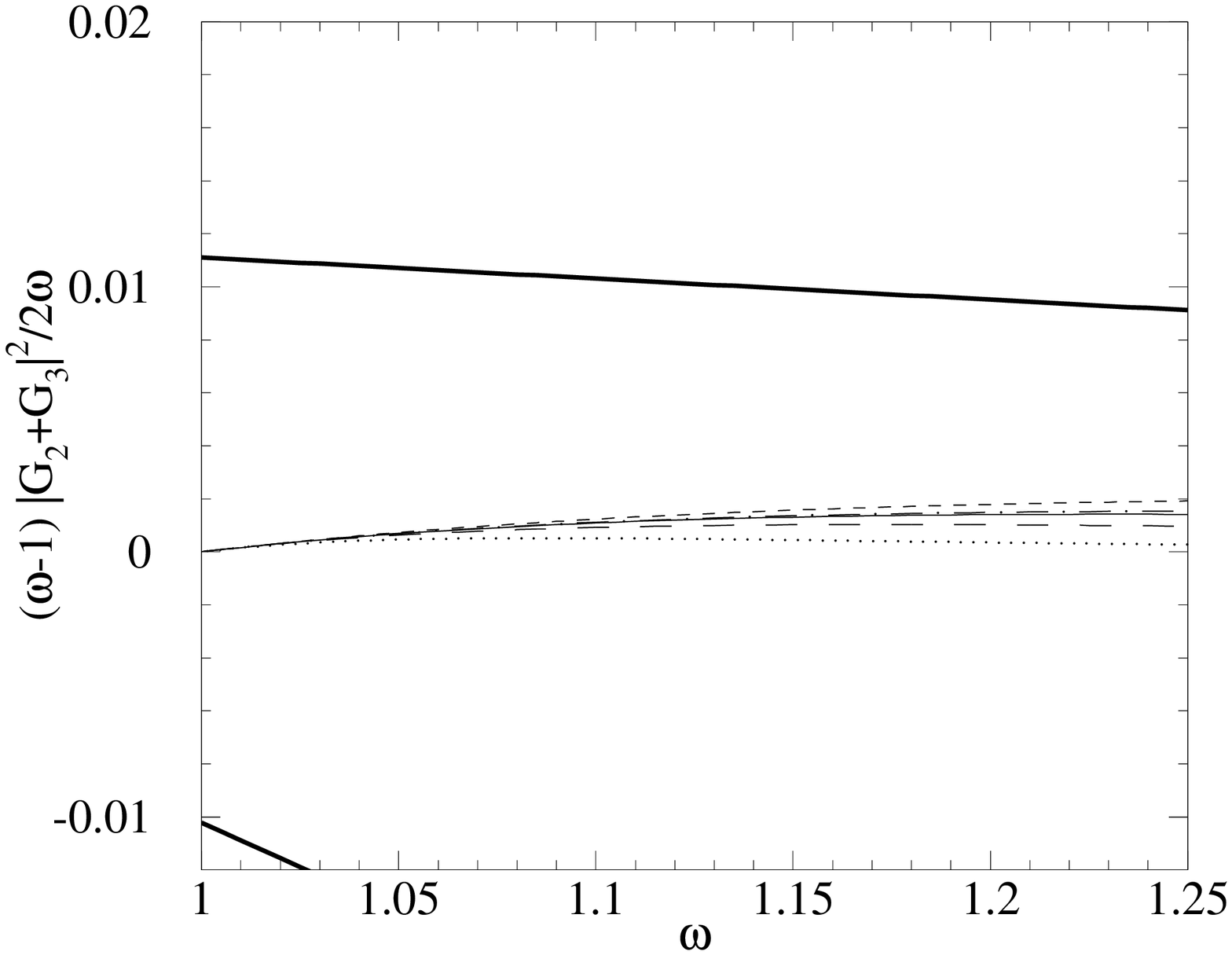} }
\tighten{
\caption[]{\it The model values of $(\omega-1)\left| G_2+G_3
\right|^2/2\omega$ along with the corresponding bounds for comparison.
The curves are labeled the same as in Fig.~5.}}
\end{figure}

The scale we chose for these plots was $\Delta = 1 {\rm\ GeV}$, since
this gives tighter bounds.  Had we chosen $2 {\rm\ GeV}$ as our
working scale, the bounds would be less stringent and thus would
accommodate the models which originally fell slightly outside our
bounds.

% ==============================================================

\section{Order $\alpha_s^2 \beta_0$ corrections at zero recoil}

As in the case of heavy mesons, one can perform analogous numerical
calculation to get the $\alpha_s^2 \beta_0$ contribution to structure
functions and the bounds at zero recoil \cite{SV94,KLWG96}.  The results for
the zeroth and first moments of the structure functions are presented
in Table~1 of \cite{CL99}, for $\Delta = 1,2 {\rm\ GeV}$.

The $O(\alpha_s^2 \beta_0)$ corrections to the upper and lower bounds
on the form factors at zero recoil are shown in Table~1 for $\Delta =
1{\rm\ GeV}$.  In this case, the $O(\alpha_s^2 \beta_0)$ corrections
are seen to be rather small, which gives us confidence that the
perturbative expansion for the bounds is under control.

\begin{table}[ht]
\begin{tabular}{c|cccc}  
& Tree Level & $O(\alpha_s)$ & $O(\alpha_s^2\beta_0)$ \\ \hline
Upper $\frac{\omega-1}{2\omega} |F_1|^2$ & 0 & 0.0063 & 0.0042 \\
Lower $\frac{\omega-1}{2\omega} |F_1|^2$ & 0 & -0.0060 & -0.0028 \\
\hline
Upper $\frac{\omega^2-1}{2\omega} \left| F_2 - F_3
\right|^2$ & 0 & 0.0127 & 0.0083 \\
Lower $\frac{\omega^2-1}{2\omega} \left| F_2 - F_3
\right|^2$ & 0 & -0.0119 & -0.0056 \\
\hline
Upper $\frac{\omega+1}{2\omega} |G_1|^2$ & 1 & -0.0713 & 0.0159 \\ 
Lower $\frac{\omega+1}{2\omega} |G_1|^2$ & 1 & -0.0885 & 0.0031 \\ 
\hline
Upper $\frac{\omega-1}{2\omega} \left| G_2 + G_3 \right|^2$
 & 0 & 0.0111 & 0.0071 \\
Lower $\frac{\omega-1}{2\omega} \left| G_2 + G_3 \right|^2$
 & 0 & -0.0102 & -0.0026 \\
\end{tabular} \vspace{6pt}
\caption{Bound on form factors at zero recoil, evaluated with
$\Delta = 1{\rm\ GeV}$.}
\end{table}

%===============================================================
\section{Bounds on Semileptonic Decay Spectrum and Its Slope}

The differential decay rate of $\Lambda_b \rightarrow \Lambda_c l \nu$ is
\begin{equation}
\label{decayrate}
\frac{{\rm d} \Gamma (\Lambda_b \rightarrow \Lambda_c l \nu)}
     {{\rm d} \omega} =
\frac{G_F^2}{24 \pi^3} |V_{cb}|^2 m_{\Lambda_b}^5 z^3
     \sqrt{\omega^2-1} \, F(z,\omega)^2, \\
\end{equation}
where
\begin{eqnarray}
\label{ratefactor}
F(z,\omega)^2 &=&
  (\omega-1) \left| (z+1) F_1 + (\omega+1) (z F_2+F_3) \right|^2 \\
&&+ (\omega+1) \left| (z+1) G_1 + (\omega-1) (z G_2+G_3) \right|^2
  \nonumber \\
&&+ 2 (z^2-2\omega z+1) \left[ (\omega-1) |F_1|^2 
                           + (\omega+1) |G_1|^2 \right]. \nonumber
\end{eqnarray}
In the linear approximation in $\omega-1$, $F(z,\omega)^2$ can be
expanded to be
\begin{equation}
\label{slopedef}
F(z,\omega)^2 = F(z,1)^2 
\left[ 1 - \rho_{\Lambda_b}^2 (\omega-1) + ... \right],
\end{equation}
where $\rho_{\Lambda_b}^2$ is the slope of the spectrum at zero
recoil.  The combination of structure functions used to bound
$F(z,\omega)^2$ is
\begin{eqnarray}
\label{ratebound}
T(\epsilon) &=& 2\omega \left[
  3 (z^2 - 2\omega \, z + 1) (T_1^{VV}+T_1^{AA})
+ 2 (\omega \, z - 1) \, \epsilon \, (T_1^{VV}+T_1^{AA}) \right. \\
&& \qquad \left.
+ \, \epsilon^2 (T_1^{VV}+T_1^{AA})
+ z^2 (w^2 - 1) (T_2^{VV}+T_2^{AA}) \right]. \nonumber
\end{eqnarray}
The bounds are drawn in Fig.~9.
\begin{figure}[t]
\centerline{\epsfysize=11truecm  \epsfbox{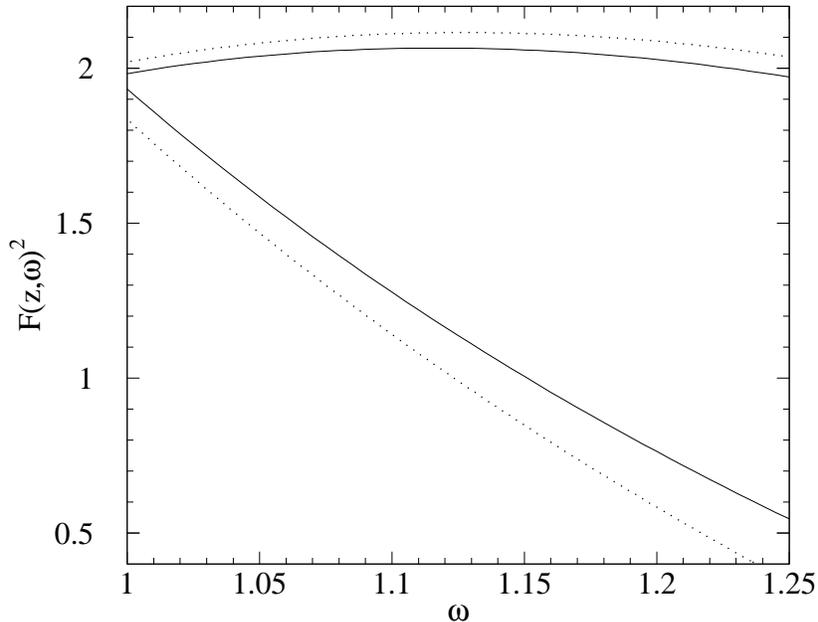} }
\tighten{
\caption[]{\it Upper and lower bounds on $F(z,\omega)^2$.  The solid
(dotted) curves are the upper and lower bounds including perturbative
corrections for HQET parameter set (A) described in the text, and
$\Delta = 1{\rm\ GeV}\ (2{\rm\ GeV})$.}}
\end{figure}
Here we use the decay rate spectrum over the full kinematic range to
extract the bounds for the slope.  As seen from the bounded region in
the plot, the allowed range of the slope is $0.87 < \rho_{\Lambda_b}^2
< 4.82$.  This agrees with the lattice calculation \cite{UKQCD97} of
$\rho_{\Lambda_b}^2 = 2.4 \pm 4$.

%===============================================================
\section{Discussion}

Many of the general features of our bounds had been mentioned in our
previous work \cite{CL99}.  In the specific application of these
bounds to heavy baryons, we find that they provide more stringent
conditions on the leading form factors, $F_1$ and $G_1$.  Looser
bounds hold for ``subleading'' form factors that are suppressed by
$1/M_Q$ in magnitude, namely, $|F_2-F_3|$ and $|G_2+G_3|$.  Our bounds
typically have much better predictive power near maximal momentum
transfer.  However, they are not stringent enough in the above
mentioned ``subleading'' form factors.  This is because both the form
factors are too small and the whole factor is suppressed by $\omega -
1$.  However, perturbative corrections do not give vanishing
contributions to the bounds at zero recoil.  The bounds also become
less stringent as $\Delta$ increases.  Therefore, we should use the
smallest value of $\Delta$ for which the perturbative expansion still
works.

We also observed that the $O(\alpha_s^2 \beta_0)$ corrections to the
bounds at zero recoil are small for $\Delta = 1{\rm\ GeV}$.  This in
turn suggests that a perturbative expansion in this problem works well
at this scale, provided that the $O(\alpha_s^2 \beta_0)$ corrections
dominate in the complete $O(\alpha_s^2)$ corrections.  (Typically
$O(\alpha_s^2 \beta_0)$ accounts for about $90\%$ of the complete
$O(\alpha_s^2)$ corrections.)

We show the bounds for the differential decay rate of $\Lambda_b
\rightarrow \Lambda_c l \nu$ and give bounds on its slope, $0.87 <
\rho_{\Lambda_b}^2 < 4.82$.  None of the models quoted here take into
account the renormalization corrections, and therefore they fail
specific bounds, such as the one in Fig.~7, as $\omega \rightarrow 1$.
In particular, this will affect the extraction of the
Kobayashi-Maskawa matrix element $|V_{cb}|$ from the zero recoil limit
of the semileptonic decay spectrum.  Also, this will give an incorrect
estimate on the experimental backgrounds and efficiencies.  Therefore,
the models should be properly modified accordingly to have more
sensible form factors.

% ==============================================================

\acknowledgments

I would like to thank Adam Leibovich in particular for discussions and
help on various technical issues.  It is also a pleasure to thank Fred
Gilman and Ira Rothstein for valuable suggestions.  This work was
supported in part by the Department of Energy under Grant
No. DE-FG02-91ER40682.

% ==============================================================

{\tighten

}

\end{document}